\documentclass{elsart}

\usepackage{amssymb}
\usepackage{amsmath}
\usepackage{amssymb}

\begin{document}

\begin{frontmatter}



\title{Reinforcement learning in market
games}


\author[eda]{Edward W. Piotrowski}
\address[eda]{Institute of Mathematics,
University of Bia\l ystok, Lipowa 41, Pl 15424 Bia\l ystok,
Poland}\ead{ep@wf.pl}
\author[jaa]{Jan S\l adkowski\corauthref{cor1}}
\address[jaa]{Institute of Physics, University of Silesia, Uniwersytecka
4, Pl 40007 Katowice, Poland}\ead{jan.sladkowski@us.edu.pl }
\author[sza]{Anna Szczypi\'nska}
\address[sza]{Institute of Physics, University of Silesia, Uniwersytecka
4, Pl 40007 Katowice, Poland}\ead{zanna11@wp.pl }
\corauth[cor1]{Corresponding author.}
\begin{abstract}
Financial markets investors  are involved in many
games -- they must interact with other agents to achieve their
goals. Among them are those directly connected with their activity
on markets but one cannot neglect other aspects that influence
human decisions and their performance as investors. Distinguishing
 all subgames is usually beyond hope and resource consuming. In
 this paper we study how investors facing many different games, gather
 information and form their decision despite being unaware of the
  complete structure of the game. To this end we apply
 reinforcement learning methods  to the Information Theory Model of Markets (ITMM).
 Following Mengel, we can try to distinguish a class $\Gamma$ of
 games and possible actions (strategies) $a^{i}_{m_{i}}$ for $i-$th
 agent. Any agent divides the whole class of games into analogy subclasses she/he thinks
 are analogous and therefore adopts the same strategy for a given
 subclass. The criteria for partitioning are based on profit and costs analysis.
 The analogy classes and strategies are updated at various stages through the process of
 learning. We will study the asymptotic behaviour of the process
 and attempt to identify its crucial stages, eg existence of
 possible fixed points or optimal strategies. Although we focus more on the instrumental aspects of agents
behaviours, various algorithm can be put forward and used for
automatic investment.  This line of research can be continued in
various directions.
\end{abstract}

\end{frontmatter}


\label{}


\vspace{5mm}
\begin{flushleft}
{\bf Motto:}
\end{flushleft}

\begin{center}
{\em "The central problem for gamblers is to find positive
expectation bets.
    But the gambler also needs to know how to manage his money, i.e. how much to bet.
    In the stock market (more inclusively, the securities markets) the problem
    is similar but more complex.
    The gambler, who is now an investor, looks for excess risk adjusted return." }\\
   \,\hfill
   \scriptsize Edward O. Thorp
\end{center}

\vspace{5mm}
\section{Introduction}
{\it Noise or structure?} We face this question almost always
while analyzing large data sets. Patern discovery is one of the
primary concerns in various fields in research, commerce and
industry. Models of optimal behaviour often belong to that class
of problems. The goal of an agent in a dynamic environment is
 to make optimal decision over time. One usually have to discard a
 vast amount of data (information) to obtain a concise model or algorithm. Therefore  prediction of individual
 agent behaviours is often burdened with large errors. The prediction game algorithm can be described as follows.

 \begin{description}
 \item FOR $n=1,2, \ldots$
    \item Reality announces $x_n \in X$
    \item Predictor announces $\gamma_n \in \Gamma$
    \item Reality announces $y_n \in Y$
\item END FOR,
 \end{description}
where $x_n \in X$ is the data upon which the prediction $\gamma_n
\in \Gamma$ is made at each round $n$. The prediction quality is
measured by some utility function $\upsilon: \Gamma \times Y
\rightarrow \mathbb{R}$.    One can
 view such a process as a communication channel that transmit
 information from the past to the future \cite{cru1}. The
 gathering of information, often indirect and incomplete, is
 referred to as measurements. Learning theory deals with the
 abilities and limitations of algorithms that learn or estimate
 functions from data. Learning helps with optimal behaviour decisions by adjusting agent's strategies
to information gathered over time. Agents can base their action
choices on prediction of the state of the environment or on reward
received during the process. For example, Markov decision process
can be formulated as a problem of finding a strategy $\pi$ that
maximizes the expected sum of discounted rewards:
$$\upsilon (s,\pi)=r(s,a_{\pi})+ \beta\sum_{s'}
p(s'|s,a_{\pi})\,\upsilon (s',\pi),$$ where $s$ is the initial
state, $a_{\pi}$ is the action induced by the strategy $\pi$, $r$
is the reward at stage  $t$ and  $\beta$ is the discount factor;
$\upsilon$ is called the value function. $p(s'|s,a_{\pi})$ denote
the (conditional)  probability\footnote{In a more formal setting
it would be a transition kernel of for the process of consecutive
actions and observations.} of reaching the state $s'$ from the
state $s$ as result of action  $a_\pi$. It can be shown that, in
the case of infinite horizon, an optimal strategy $\pi^{\ast}$
such that (Bellman optimality equation)
$$\upsilon (s,\pi^{\ast})=\max_{a} \{r(s,a)+ \beta\sum_{s'}
p(s'|s,a)\,\upsilon (s',\pi ^{\ast})\}$$ exists. In reinforcement
learning, the agent receives rewards from the environment and uses
them as feedback for its action. Reinforcement learning has its
roots in statistics cybernetics, psychology, neuroscience,
computer science ... .  In its standard formulation, the agent
must improve his/her performance in a game through trial-and-error
interaction with a  dynamical environment. There are two ways of
finding the optimal strategy:\begin{description}
    \item {\it strategy iteration} -- one directly manipulates the
    strategy;
    \item {\it value iteration} -- one approximates the optimal value
    function.
\end{description}
Therefore two classes of algorithms are considered: strategy
(policy) iteration algorithms and value iteration algorithms.  In
the following section we discuss the adequacy of reinforced
learning in market games.
\section{Reinforcement learning in market games}
 Can reinforcement learning  help with market games analysis?
 Could it be used for finding optimal strategies? It not easy to
 answer this question because it involves the problem of real-time
 decision making one often have to (re-)act as quickly as possible.
 Consider model-free reinforcement learning,
 Q-learning\footnote{This is obviously a value iteration, but in
 market games there is a natural value function -- the profit.} In
 this approach one defines the value of an action $Q(s,a)$ as a
 discounted return if action $a$ following from the strategy $\pi$
 is applied:
 $$ Q^{\ast}(s,a)= r(s,a) + \beta \sum_{s'}
p(s'|s,a)\,\upsilon (s',\pi ^{\ast})$$ then
$$\upsilon (s,\pi^{\ast})=\max_{a} Q^{\ast}(s,a)\,.$$
In Q-learning, the agent starts with arbitrary $Q(s,a)$ and at
each stage $t$ observes the reward $r_{t}$ and the updates  the
value of  $Q$ according to the rule: $$ Q_{t+1}(s,a)=(1-\alpha
_{t})\,Q_{t}(s,a) + \alpha _{t}(r_{t} +\beta\max_{b}Q_{t}(s,b)),
$$
where  $\alpha _{t}\in [0,1)$ is the learning rate that needs to
 decay over time for the learning algorithm to converge. This
 approach is frequently used in stochastic games setting. Watkins and Dayan proved that this sequence converges
provided all states and  actions have been visited/performed
infinitely often \cite{WD}. Therefore we anticipate weak
convergence ratios. Indeed, various  theoretical and experimental
analyses \cite{URL,RCP,Littman} showed that even in very simple
games might require $\sim 10^8$ steps! If  a well-shaped stock
trend is formed, one can expect that there are sorts of {\it
adversarial equilibria} (no agent is hurt by any change of others'
strategies)
$$ R_i(\pi _1,\ldots ,\pi_n) \leq R_i(\pi_1',\ldots
,\pi'_{i-1},\pi'_{i},\pi'_{i+1},\ldots ,\pi_n')$$  or {\it
coordination equilibria} (all agents achieve their highest
possible return)
$$ R_i(\pi _1,\ldots
,\pi_n) =\max _{a_1,\ldots , a_n} R_i(a_1,\ldots ,a_n).$$ Here
$R$s denote the pay-off functions and $\pi$s the one-stage
strategies. The problem is they can be easily identified with
technical analysis\footnote{We understand the term technical
analysis as simplified hypothesis testing methods that can be
applied in real time.} tools and there is no need to recall to
learning algorithms. In the most interesting classes of games
neither adversarial equilibria nor coordination equilibria exist.
This type of learning is much more subtle and, up to now, there is
no satisfactory analysis in the field of reinforcement learning.
Therefore a compromise is needed, for example we must be willing
to accept returns that might not be optimal. The models discussed
in the following subsections belong to that class and seem to be
tractable by leaning algorithms.
\subsection{Kelly's criterion}
Kelly's criterion \cite{kelly} can be successfully applied in
horse betting or blackjack when one can discern biases
\cite{Thorpe} even though its optimality and convergence   can be
proven only in the asymptotic cases. The simplest form of Kelly's
formula is:
$$ \Theta= W - (1-W)/R$$ where:
\begin{itemize}
  \item $\Theta =$ percentage of capital to be put into a single trade.
  \item $W = $ historical winning percentage of a trading system.
  \item $R =$ historical average win/loss ratio.
\end{itemize}
Originally, Kelly's formula involves finding the
     "bias ratio" in a biased game. If the game is
     infinitely often repeated then one should put at stake
     the percentage of one's capital equal to the bias ratio.
     Therefore one can easily construct various learning
     algorithms that perform the task of finding an environment so that Kelly's approach can be
effectively applied (bias search + horizon of the investment)
\cite{schroe,PL}.
\subsection{MMM model}

 Piotrowski and S\l adkowski have analysed the model where the trader fixes a maximal price
he is willing to pay for the asset $\Theta $ and then, if the
asset is bought, after some time sells it at random \cite{MMM}.
One can easily reverse the buying and selling strategies. The
expected value of the of the profit after the whole cycle is

    $$ \rho _{\eta}\left( a \right)= \frac{-\int
_{-\infty}^{-a}p\;\eta \left( p\right) dp}{1+ \int
_{-\infty}^{-a}\eta \left( p\right) dp}\, $$ where $a$ is the withdrawal
price. The maximal value of the function $\rho$, $a_{max}$, lies
at a fixed point of $\rho$, that is fulfills the condition $\rho
\left( a_{max}\right) = a_{max}$. The simplest version of of the
strategy is as follows: {\it there an optimal strategy that fixes
the withdrawal price at the level historical average
profit\footnote{ Or else: do not try to outperform yourself.}}.
   Task: find an implementation of reinforced learning algorithm that can be used effectively on
   markets. We should control both, the probability distribution $\eta$ and the profit "quality".

\subsection{Learning across games} An interesting approach was put
forward by Mengel \cite{Men}. One can easily give examples of
situations where agents cannot be sure in what game they are
taking part (e.g. the games  may have the same set of actions).
Distinguishing all possible games and learning separately for all
of them requires a large amount of alertness, time and resources
(costs). Therefore the agent should try to identify some classes
of situations she/he perceives as analogous and therefore takes
the same actions. The learning algorithm should update both the
partition of the set of games and actions to be taken:
\begin{itemize}
    \item Agents are playing repeatedly a game (randomly) drawn from a set $\Gamma$
   \item Agents partition the set of all games into subset (classes) of games they
   learn not to discriminate (see them as analogous)
    \item Agents update both  propensities to use partitions \{G\} and
    attractions towards using their possible strategies/actions
\end{itemize}
 Asymptotic behaviour and computation complexity of such process is discussed in Ref. \cite{Men}.
 Stochastic approximation is working in this case (approximation through a system of deterministic
  differential equations is possible). It would be interesting to analyse the following problems.   Problem 1.: Identify possible "classes of market  games"
Problem 2.: Identify "universal" set of strategies. For example,
on the stock exchange one can try the brute force approach. Admit
as strategies  buying/selling at all possible price levels and
identify classes of games with trends. Unfortunately, the number
of approximations generates huge transaction costs. This can be
reduced on the derivative markets as due to the leverage the ratio
of transaction cost to price movements is much lower. We envisage
that an agent may try to optimize among various classes of
technical analysis tools.

\section{Conclusion}
As conclusions we would like to stress the following points.
\begin{description}

 \item Algorithms are simple but computation is complex, time and  resource
 consuming.
    \item Learning across games could be used to "fit" technical analysis
    models.
  \item Dynamic proportional investing (Kelly) should be the easiest to
  implement. But here we envisage problems analogous to related to
  heat (entropy) in thermodynamics, and exploration of knowledge
  might involve in cases of high effectiveness paradoxes
  \cite{schroe} analogous to those of arising when speed
  approaches the speed of light \cite{PL}.
\item One can envisage learning (information) models of markets/
portfolio theory. \item Implementation should be carefully tested
-- transaction costs can "kill" even crafty algorithms
\cite{arbitrage}. \item Quantum algorithms/computers, if ever come
true might
  change the situation in a dramatic way: we would have powerful
  algorithms at our disposal and and the learning limits would
  certainly  broaden \cite{bali,q-comp,qai}.
\end{description}


\begin{thebibliography}{00}

 \bibitem{cru1}  Crutchfield, J. P., Shalizi, C. R., {\it  Physical
 Review E} {\bf 59} (2003) 275.
\bibitem{benv} Benveniste A., M. Metevier and P. Priouret (1990), Adaptive Algorithms
and Stochastic Approximation, Berlin: Springer Verlag.
\bibitem{fuden} Fudenberg, D. and D.K. Levine (1998), The Theory of Learning in Games,
Cambridge: MIT-Press.
\bibitem{kusch} Kuschner, H.J. and G.G. Lin (2003), Stochastic Approximation and Recursive
Algorithms and Applications, New York: Springer.
\bibitem{WD} Watkins, C.J.C.H., Dayan, P., Q-learning, {\it
Machine Learning} {\bf 8} (1992) 279.
\bibitem{URL} Farias, V.F., Moallemi,C.C.,  Weissman, T.,  Van Roy,B., Universal Reinforcement Learning, arXiv:0707.3087v1
[cs.IT].
\bibitem{RCP} Shneyerov, A., Chi leung Wong, A., The rate of convergence to a perfect
competition of a simple Matching and bargaining Mechanism,Univ of Britich
Columbia preprint (2007).
\bibitem{Littman} Littman, M.L., Markov games as a framework for
multi-agent reinforcement learning, Brown Univ. preprint.
\bibitem{kelly} Kelly, J.L., Jr.,
A New Interpretation of Information Rate, {\it The Bell System
Technical Journal} {\bf 35} (1956)
 917.
\bibitem{Thorpe} Thorpe, E.O., Optimal gambling systems for favorable games, {\it Revue de
l'Institut International de Statistique / Review of the
International Statistical Institute}, {\bf 37} (1969) 273.
\bibitem{schroe} Piotrowski, E.~W., Kelly Criterion
revisited: optimal bets, {\it Physica A\/}  (2007), in press.
\bibitem{PL} Piotrowski, E.~W., \L uczka, J., The relativistic
velocity addition law optimizes a forecast gambler's profit,
submmited to {\it Physica A\/}; arXiv:0709.4137v1
[physics.data-an].
\bibitem{MMM} Piotrowski, E.~W., S\l adkowski, J., The Merchandising Mathematician Model,
    {\it Physica A\/} {\bf 318} (2003) 496.
\bibitem{Men} Mengel, F., Learning across games, Univ. of Alicante report WP-AD
2007-05.
\bibitem{arbitrage} Piotrowski, E.~W., S\l adkowski, J., Arbitrage risk induced by transaction
costs, {\it Physica A\/} {\bf 331} (2004) 233.
\bibitem{bali} Piotrowski, E.~W., Fixed point theorem for simple quantum strategies in quantum
market games, {\it Physica A\/} {\bf 324} (20043) 196.
\bibitem{q-comp} Piotrowski, E.~W., S\l adkowski, J., Quantum computer: an appliance for playing market
games, {\it International Journal of Quantum Information} {\bf 2}
(2004) 495.
\bibitem{qai} Miakisz, K.,Piotrowski, E.~W., S\l adkowski, J., Quantization of Games: Towards Quantum Artificial
      Intelligence, {\it Theoretical Computer Science} {\bf 358} (2006) 15.
\end{thebibliography}
\end{document}